\documentclass[twocolumn]{emulateapj}  
\usepackage{epstopdf}
\usepackage{ulem}
\usepackage{amssymb}
\usepackage{natbib}
\usepackage{times}
\usepackage{graphicx,bm,amssymb}
\usepackage{pstricks}

\usepackage{psfrag}
\usepackage[colorlinks=true,linkcolor=blue,citecolor=blue]{hyperref}
\def\aj{AJ}
\def\apj{ApJ}
\def\apjl{ApJ}
\def\apjs{ApJS}
\def\aap{A\&A}
\def\aaps{A\&AS}
\def\mnras{MNRAS}
\def\prd{Phys.~Rev.~D}
\def\physrep{Phys.~Rep.}

\newcommand{\be}{\begin{equation}}
\newcommand{\ee}{\end{equation}}
\newcommand{\bse}{\begin{subequations}}
\newcommand{\ese}{\end{subequations}}
\newcommand{\bary}{\begin{eqnarray}}
\newcommand{\eary}{\end{eqnarray}}
\newcommand{\en}{E_\nu}

\slugcomment{Submitted to ApJ}

\shorttitle{The GeV counterpart of VER J2019+407}
\shortauthors{N. Fraija and M. Araya}
\begin{document}
\title{The GeV counterpart of VER J2019+407 in the northern shell of the supernova remnant G78.2+2.1 ($\gamma$ Cygni)}
\title{The GeV counterpart of VER J2019+407 in the northern shell of the supernova remnant G78.2+2.1 ($\gamma$ Cygni)}
\author{N. Fraija\altaffilmark{1} and M. Araya\altaffilmark{2} }
\affil{$^1$Instituto de Astronom\'ia, Universidad Nacional Aut\'{o}noma de M\'{e}xico, Apdo. Postal 70-264, Cd. Universitaria, DF 04510, M\'{e}xico}
\affil{$^2$Escuela de F\'isica \& Centro de Investigaciones Espaciales (CINESPA), Universidad de Costa Rica, San Jos\'e 2060, Costa Rica}
\email{nifraija@astro.unam.mx and miguel.araya@ucr.ac.cr}
\date{\today}
\begin{abstract}
Analysis of gamma-ray emission from the supernova remnant G78.2+2.1 ($\gamma$ Cygni) with 7.2 years of cumulative data from the Fermi-LAT telescope shows a distinct hard, bright and extended component to the north of the shell coincident with the known TeV source VER J2019+407.   In the GeV-TeV energy range its spectrum is best described by a broken power-law with indices 1.8 below a break energy of 71 GeV and 2.5 above the break.  A broadband spectral energy distribution is assembled and different scenarios for the origin of the gamma-rays are explored. Both hadronic and leptonic mechanisms are able to account for the GeV-TeV observations. In the leptonic framework, a superposition of inverse Compton and nonthermal bremsstrahlung emissions is needed whereas the hadronic scenario requires a cosmic ray population described by a broken power-law distribution with a relatively hard spectral index of $\sim1.8$ below a break particle energy of 0.45 TeV.  In addition, the neutrino flux expected from cosmic ray interactions is calculated.
\end{abstract}
\section{Introduction}
Supernova remnants (SNRs) are believed to be the sites where cosmic rays (CR) in our galaxy are accelerated up to $\sim$ PeV energies. Studying the GeV-TeV $\gamma$-ray emission from SNRs is important to shed light on the puzzle of the origin of these high energy (HE) particles which populate the Galaxy. The mechanism of diffusive shock acceleration \cite[e.g.,][]{1978MNRAS.182..147B,1987PhR...154....1B} is thought to play an important role in transferring some of the kinetic energy from a SNR's expansion to particles. The shocks of SNRs are known to accelerate electrons up to very high energies (VHEs). In the leptonic framework, electron synchrotron radiation is emitted from radio wavelengths to X-rays \cite[e.g.,][]{1994AstL...20..157A,2001ApJ...552L..39G,2002A&A...395..943B,2002ApJ...581.1101H,2002ApJ...581.1116R}, and inverse Compton scattering  and bremsstrahlung emissions are emitted in GeV $\gamma$-rays.  In the hadronic picture, CR protons (or nuclei) can produce GeV-TeV $\gamma$-rays through the decay of neutral pions produced in photo-hadronic interactions as well as hadronic collisions with ambient material. Therefore, spectral and morphological studies of $\gamma$-ray data can be used in principle to disentangle the nature of the particles involved. 
Although, during the last years, many SNRs have been detected in $\gamma$-rays \cite[e.g.,][]{2009ApJ...706L...1A,2010ApJ...722.1303A,2010ApJ...718..348A,2010ApJ...710L..92A,2010ApJ...712..459A,2010Sci...327.1103A,2010ApJ...712..287E,2011ApJ...735..120Y,2011ApJ...740L..51T,2012ApJ...744...80A, 2011Sci...334.1103A, 2013MNRAS.434.2202A,2014MNRAS.444..860A,2015ApJ...813....3A}, it is not clear which emission mechanism dominates in most cases. In some SNRs the $\gamma$-ray emission is coincident with high density ambient clouds and its spectral shape can be described by the hadronic scenario, although details on the predicted spectral shape are still debated and the ambient properties are usually poorly known.\\
Some related TeV sources are pulsar wind nebulae (PWN), the relativistic magnetized winds (mostly $e^\pm$ pairs, associated to a pulsar) that create termination shocks where particle acceleration occurs. Their spectral energy distributions (SEDs) are dominated by synchrotron and inverse Compton scattering from HE leptons, although hadronic processes may also contribute \citep{1990JPhG...16.1115C,2003A&A...402..827A}. On the other hand, however, the nature of many of the VHE sources remains unknown. Such is the case of the source VER J2019+407 \citep{2013ApJ...770...93A}\footnote{See also the catalog of TeV sources http:\/\/tevcat.uchicago.edu}. At TeV energies this source has an extent of $0^{\circ}.23\pm0^{\circ}.03_{\mbox{\tiny stat}}\,^{+0^{\circ}.04}_{-0^{\circ}.02\mbox{\tiny sys}}$ and its $\gamma$-ray spectrum is described by a simple power law \citep{2013ApJ...770...93A}.\\
VER J2019+407 is found within the radio shell of the SNR G78.2+2.1 in spatial coincidence with non thermal radio emission \citep{2008A&A...490..197L}. Located at (1.5 - 1.8) kpc and seen at $\sim$2.5 degrees from the star forming region Cygnus X \cite[e.g.,][]{1991A&A...241..551W}, it is $\sim1^{\circ}$ in diameter \citep{1977AJ.....82..329H,1977AJ.....82..718H,1980A&AS...39..133L,2000AstL...26...77L}. The SNR shows a distinctive shell in radio wavelengths and X-rays with brighter northern and southern regions \citep{2013MNRAS.436..968L,2006A&A...457.1081K,1997A&A...324..641Z,1996MNRAS.281.1033B,1977AJ.....82..329H,1966ApJ...144..937D}. Most of the radio flux comes from the southeastern part of the remnant. Higher radio spectral indices (steeper spectra) are seen near the center of the remnant (although they could be associated to the shell), and around the shock of the SNR \citep{1997A&A...324..641Z}. Several studies have found that the average radio spectral index is around 0.5 \cite[e.g.,][]{1988A&A...191..313P,2006A&A...457.1081K}, but a detailed analysis of the ambient thermal emission and the synchrotron radiation of the remnant with high sensitivity data from the Canadian Galactic Plane Survey \cite[CGPS;][]{2003AJ....125.3145T} has yielded a significantly larger average spectral index of $0.75\pm0.3$ with spatial variations from 0.40 to 0.80 \citep{2008A&A...490..197L}. There is evidence that the non thermal radio emission at the location of the TeV source VER J2019+407 has a spectrum that is harder than average, with a spectral index close to 0.40 \citep{1997A&A...324..641Z}.\\
Neutral hydrogen emission has been observed and associated with the SNR including post-shock HI in the south and north-east of the SNR \citep{1980A&AS...39..133L,1986A&AS...63..345B} as well as a shell around the SNR with an estimated density of 2.5 cm$^{-3}$ that might have been produced by the stellar wind of the SNR progenitor \citep{2001AstL...27..233G}. No evidence has been found of interactions with molecular gas in a CO survey \citep{1983AJ.....88...97H}. However, \citet{1988srim.conf..261F} observed a CO cloud which was associated to the south-eastern section of the SNR and \cite{1988A&A...191..313P} found H$_2$CO absorption, indicating the presence of dense material in this zone. A large thermal radio feature and several small HII regions are imaged near and overlapping parts of the SNR, among them the HII region known as the $\gamma$ Cygni nebula coincident in position with the brightest non-thermal emission in the south-east of the SNR \citep{1977AJ.....82..329H,1991A&A...241..551W}.\\
Requiring a distance of 1.7 kpc, X-ray and H$\alpha$ observations have yielded a shock velocity of $(1-1.5)\times10^3$ km s$^{-1}$,  an unperturbed number density $(0.14-0.3)$ cm$^{-3}$, a temperature of $(1.6-3.2)\times10^7$ K for the X-ray emitting gas in the bright shell and a remnant age of 4000 to 6000 yr, assuming an explosion kinetic energy of $10^{51}$ erg \citep{2000AstL...26...77L}. Analyzing optical data of bright regions, \citet{2003A&A...408..237M} found a shock speed of 750 km s$^{-1}$, an age of 7000 yr and unperturbed ISM densities of 0.3 cm$^{-3}$ containing compressed clouds with densities of $\sim700$ cm$^{-3}$ (corresponding to pre-shock densities of $\sim20$ cm$^{-3}$). Based on ASCA X-ray observations, Uchiyama et al. \citep{2002ApJ...571..866U} derived a shock velocity of 800 km s$^{-1}$ and an adiabatic age of 6600 yr, assuming a distance of 1.5 kpc. These authors also discovered two clumps (labeled C1 and C2) of hard X-ray emission in the north of the SNR.  These  were attributed to non thermal bremsstrahlung radiation from electrons in shocked dense cloudlets with gas density between 10 and 100 cm$^{-3}$. An analysis of INTEGRAL-ISGRI \citep{2004A&A...427L..21B} revealed hard X-ray clumps of about ten arcmin in size in the north-west (the most prominent one), the south-east and the north-east regions of the SNR. These are compatible with both the ASCA fluxes derived by \cite{2002ApJ...571..866U} and the spatially unresolved RXTE PCA data \citep{2004A&A...427L..21B}. The observed spatial morphology and spectra are consistent with the emission generated by nonthermal electrons accelerated at the shock of a supernova interacting with an interstellar cloud. Using a Chandra observation of the region \citep{2013MNRAS.436..968L}, it was found later that C1 is extended and likely associated to an extragalactic object, whereas C2 has an absorption column density which is in agreement with the location of the SNR and a nonthermal, power-law, spectrum with index of $\sim1.0$ \citep{2013MNRAS.436..968L}. C2 is located within the extent of VER J2019+407.\\
Gamma-ray emission in the region of G78.2+2.1 was detected by the EGRET instrument on board the Compton Gamma Ray Observatory (CGRO) \citep{1995ApJS..101..259T}. The Large Area Telescope (LAT) on board Fermi Gamma-ray Space Telescope satellite has found extended emission from this remnant \citep{2012ApJ...756....5L} and also discovered a bright $\gamma$-ray pulsar, PSR J2021+4026, near the center of the remnant \citep{2010ApJS..187..460A,2010ApJS..188..405A}.  G78.2+2.1 is also near the Cygnus Cocoon, an extended region of freshly accelerated cosmic rays seen by the Fermi satellite above 1 GeV \citep{2011Sci...334.1103A} which also shows TeV emission \citep{2007ApJ...658L..33A, 2014ApJ...790..152B}. The SNR could be a source of at least some of the high energy particles in the cocoon \citep{2011Sci...334.1103A, 2007ApJ...658L..33A}.\\
Although the $\gamma$-ray emissions from both the SNR and the pulsar PSR J2021+4026 are difficult to disentangle at a few GeVs, the flux from the SNR becomes dominant above $\sim10$ GeV, as the spectrum of PSR J2021+4026 follows a power law with an exponential cutoff with a cutoff energy of $\sim2.4$ GeV. In the latest LAT catalog the morphology of the emission from the entire shell of the SNR, also called 3FGL J2021.0+4031e  \citep{2015ApJS..218...23A}, has been described with a uniform disk of radius 0$^{\circ}$.63 centered at the position (J2000) $\alpha = 305^\circ.27$, $\delta = 40^\circ.52$. In this paper, we report the discovery of an extended region of enhanced GeV emission that is spatially coincident with the TeV source VER J2019+407 using LAT data. The paper is arranged as follows: in Section  \ref{LATdata} we describe the data analysis and the results in Section \ref{results}. In Section \ref{model} we present the calculations of the models for the nonthermal emission and the expected neutrino flux. Our discussion is given in Section \ref{discussion} and the conclusions in Section \ref{conclusion}. We hereafter use k=$\hbar$=c=1 in natural units  and adopt a source distance of 1.5 kpc.
%
\section{Fermi LAT observations}\label{LATdata}
LAT data selected in the energy range 4-300 GeV and acquired from the beginning of the mission, 2008 August to 2015 November, are analyzed with the most recent software {\small SCIENCETOOLS} version v10r0p5\footnote{See \url{http://fermi.gsfc.nasa.gov/ssc}} and reprocessed with ``Pass 8'' photon and spacecraft data \citep{2013arXiv1303.3514A} and the instrument response functions (IRFs) P8R2\_SOURCE\_V6P.\\
Photons are selected within a square region of $20^{\circ}\times20^{\circ}$ centered at the coordinates (J2000) $\alpha$ = 305$^\circ$.34, $\delta$ = 40$^\circ$.43 (the region of interest, hereafter ROI). As indicated above, the analysis is restricted to energies above 4 GeV in order to avoid the bright emission from the pulsar PSR J2021+4026. Standard cuts are applied in the analysis, including selecting zenith angles less than 90$^{\circ}$ to remove photons from the Earth limb, and keeping the SOURCE class events. Time intervals are selected when the LAT instrument was in science operations mode and the data quality was good. The data is binned in 20 logarithmically spaced bins in energy and a spatial binning of 0$^{\circ}$.02 per pixel is used.\\
The analysis of LAT data is performed by a method of maximum likelihood which estimates the probability of reproducing the spectral and morphological properties of sources with a given model \citep{1996ApJ...461..396M}. The model initially contains the sources present in the latest LAT catalog \citep{2015ApJS..218...23A} as well as the standard diffuse Galactic emission template and the isotropic emission that accounts for extragalactic background and misclassified cosmic rays. In all the likelihood fits, the normalizations of these components are kept free as well as the normalizations of sources located less than six degrees from the center of the ROI, except for those sources which are detected below the $5\sigma$ level above 4 GeV, in which case their spectral parameters are kept fixed to the cataloged values. The effect of freeing the normalization of PSR J2021+4026 spectrum inside the shell of G78.2+2.1 is evaluated.\\
\section{Results}\label{results}
\begin{figure}
\epsscale{1.5}
\plotone{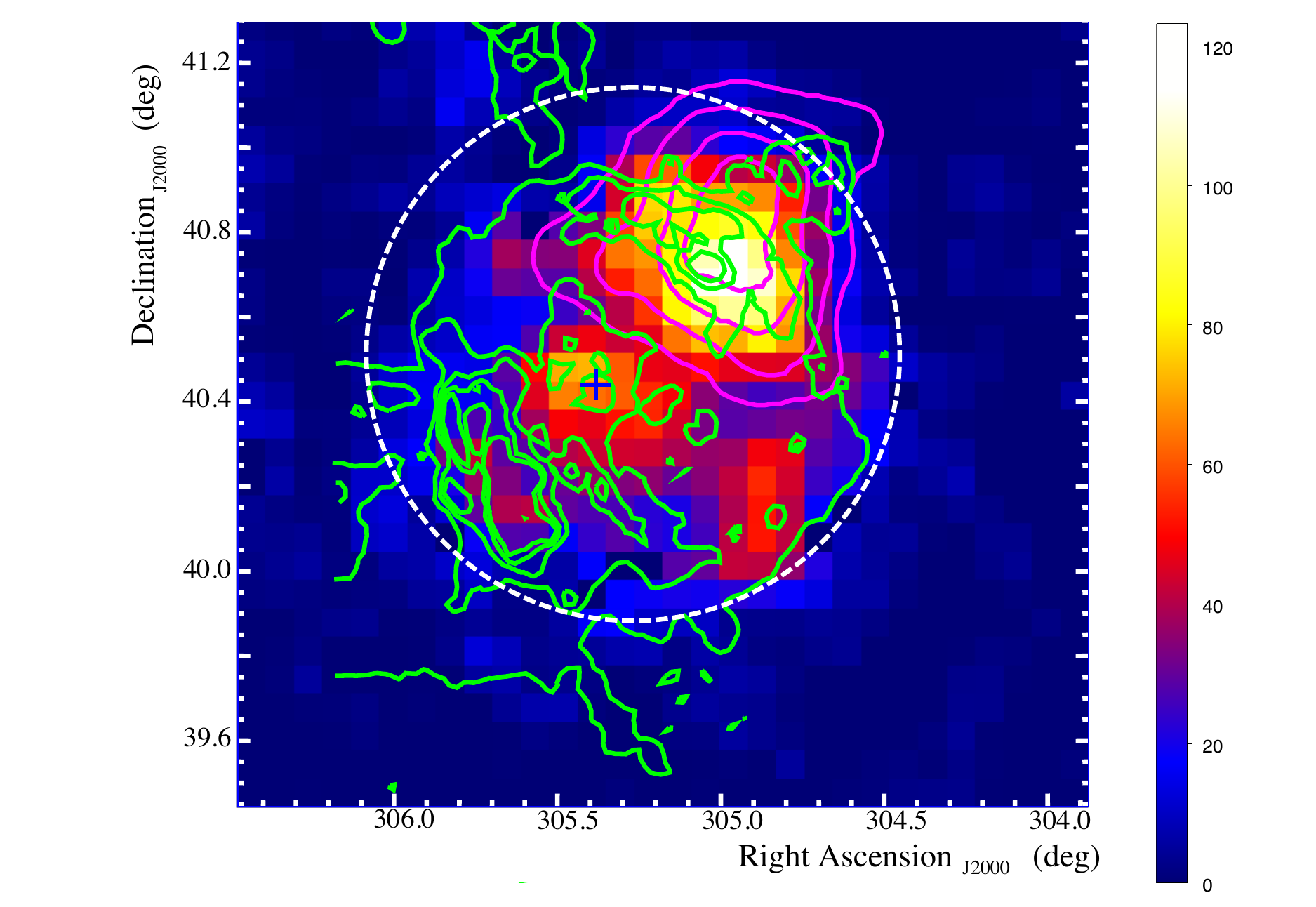}
\caption{Significance map, in units of TS, of the  G78.2+2.1 region for $\gamma$-rays above 15 GeV with a scale of $\sim0.067$ deg/pixel. The CGPS 1420 MHz observation is represented by green contours at brightness temperatures of 22 K, 31.5 K, 41 K, 50.5 K and 60 K. The VER J2019+407 smoothed photon excess contours (100, 150, 210 and 260 photons) are shown in magenta. The location of the $\gamma$-ray pulsar PSR J2021+4026 is marked with a blue cross and the boundary of the 3FGL J2021.0+4031e disc is displayed by the white dashed circle.\label{fig1}}
\end{figure}

From the LAT data analysis described above we conclude that a uniform disc is not a good description of the $\gamma$-ray morphology of G78.2+2.1. If the source 3FGL J2021.0+4031e is not included in the model the resulting residuals show a morphology that is quite unlike that of a uniform disc, which was reported in the LAT catalog for this source. Even when we incorporate this source in the model, the resulting residuals show a clear excess in the northern region of the SNR. To study this emission further, a significance map is calculated above 15 GeV with the tool gttsmap for a square region of  $2^{\circ}\times2^{\circ}$ around the position of the SNR. The map, shown in Figure \ref{fig1}, is in units of test statistic (TS) which is defined as twice the logarithm of the ratio of maximum likelihood values obtained with a point source located at each pixel position and that obtained with no source. The associated significance at each position is then $\sim \sqrt{\mbox{TS}}$ \citep{1996ApJ...461..396M}. The TS map in Fig. \ref{fig1} shows the radio contours from a CGPS observation of G78.2+2.1 as well as the significance contours of the TeV source VER J2019+407 \citep{2013ApJ...770...93A}. The same region of enhanced emission seen in the residuals map is clearly visible in the TS map, which coincides with the location of VER J2019+407.\\
The excess found in the northern shell of the SNR is extended, as shown by residuals obtained after placing a point source at the position of maximum TS in the significance map. Due to the previous result and the location of the excess in coincidence with VER J2019+407, it is natural to associate the GeV with the TeV source. We adopt the spatial morphology used to describe the TeV emission from VER J2019+407 \citep{2013ApJ...770...93A} as a template for the emission detected by the LAT telescope, which is a symmetric two-dimensional Gaussian with an extension of $0^{\circ}.23$ and centroid coordinates (J2000) $\alpha$ = 305$^0$.02, $\delta$ = 40$^0$.76.\\
The rest of the emission from the SNR around the position of VER J2019+407 has to be taken into account in the model to accurately describe the properties of the source. We then use a spatial template obtained from removing a $0^{\circ}.23$-radius circular region around VER J2019+407 from the disc of 3FGL J2021.0+4031e, to represent the southern part of $\gamma$ Cygni and the rest of the SNR's shell. We fit the data above 4 GeV adopting power law spectra for both components with free parameters. The resulting TS values for VER J2019+407 and the southern region are 180 and 400, respectively.  The power law indices and fluxes are $1.91\pm0.10$ and $(7.61\pm0.84)\times10^{-10}$ ph cm$^{-2}$ s$^{-1}$ for VER J2019+407, and $1.93\pm0.06$ and $(2.47\pm0.17)\times10^{-9}$ ph cm$^{-2}$ s$^{-1}$ for the southern region. It is important to note that these parameters are derived for events in the 4-300 GeV band, while the TS map plotted in Fig. \ref{fig1} was obtained above 15 GeV in order to avoid contamination from the bright pulsar near the center of the SNR. Even above 10 GeV, the pulsar is detected at almost the $11\sigma$ level. The determination of the spectral parameters of the southern shell is affected by the pulsar emission. It is found that freeing the spectral normalization of the pulsar model in the fit results in a spectral index for the southern shell of $2.04\pm0.06$ without considerably affecting the results for VER J2019+407. The study of the effect of the pulsar emission in the region beyond the extension of VER J2019+407 is outside of the scope of this paper and thus we cannot conclude if the spectrum of VER J2019+407 is harder than that of the rest of the shell around several GeV. However, VER J2019+407 seems to show a harder  LAT spectrum above $\sim10$ GeV. In count maps of increasing energy the emission of the source certainly dominates over the rest of the shell, which is also consistent with the non detection of TeV photons around this region \citep{2013ApJ...770...93A},  and, in the 10-300 GeV (15-300 GeV) range, the spectral indices for VER J2019+407 and the rest of the shell are $2.07\pm 0.17$ and $2.20\pm 0.10$ ($2.1\pm 0.2$ and $2.35\pm 0.15$), respectively.\\
The LAT spectrum of VER J2019+407 is not curved in the energy range analyzed here. Assuming either a log parabola or a power law with an exponential cutoff for the spectrum results in the same source significance as the one obtained with a simple power law.\\
In order to construct the LAT SED, we binned the data in 16 logarithmically spaced energy intervals from 4 to 300 GeV and also applied a likelihood fit in each interval. When the TS of the source falls below nine, a 95\% confidence level upper limit on the flux is calculated in the corresponding interval. In each bin, the spectral normalization of the pulsar is set to free to compare results when it is kept frozen to the cataloged value. The effect of the pulsar on VER J2019+407 is only important in the first energy bin (4-5.5 GeV), where it results in a 10\% variation of the flux from the source. This is considered as a systematic error in this bin.\\
A careful analysis of the LAT observation to decrease the effect of the contamination from photons from the bright nearby pulsar PSR J2021+4026, for example using ``front'' events with an improved PSF, which is beyond the scope of this work, is needed to study the emission of the SNR below a few GeV. We carried out a preliminary fit to the LAT data only, above 1 GeV, and found an indication that the photon spectrum becomes softer at lower energies. Using the same spatial template for VER J2019+407 the fit yields a power-law spectrum with index $2.22\pm0.08_{\mbox{\tiny stat}}$.
\subsection{Multiwavelength spectral energy distribution}
We gathered data from the literature to assemble the multiwavelength SED. The radio fluxes are used for the integrated emission from the entire SNR \citep{1977AJ.....82..718H,1991A&A...241..551W,1997A&A...324..641Z, 2006A&A...457.1081K,2011A&A...529A.159G} for reference. The 10.6 GHz flux from a $7'\times9'$ region within VER J2019+407 is taken from a previous study \citep{1977AJ.....82..329H} as the expected radio synchrotron flux from the northern shell of G78.2+2.1. The X-ray data are taken from RXTE PCA and ASCA observations of the entire SNR \citep{2004A&A...427L..21B} and the ASCA fluxes from the hard X-ray clumps from \cite{2002ApJ...571..866U}. The INTEGRAL-ISGRI fluxes from another clump of hard emission to the north-west of the SNR \citep{2004A&A...427L..21B} are also shown.\\
\section{Emission Model}\label{model}
The emission models and particle distributions used to account for the broadband SED of VER J2019+407 are summarized next. We use a simple one-zone model that includes standard radiation mechanisms from leptons and hadrons. In all the broadband models the emitting region is taken as spherical with a radius of $0^{\circ}.23$ which at a distance of 1.5 kpc corresponds to a physical radius of 6 pc.
\subsection{Synchrotron radiation}
In the relativistic case, the electron population is given by $N(E_e)\propto E_e^{-\alpha_e}$ where $\alpha_e$ is the spectral index and $E_e$ the particle energy. We assume this distribution cuts off above certain energy. The electrons in the emitting region permeated by a magnetic field $B$ cool down in a time scale given by the standard cooling time $\tau_c=\frac{6\pi m^2_e}{\sigma_T}\,B^{-2}E^{-1}_e$, with  $\sigma_T=6.65 \times 10^{-25}\, {\rm cm^2}$ the Compton cross section and $m_e$ the electron mass. Equating the synchrotron time scale with the age of the source $\tau_{\rm age}$ \citep{2002ApJ...571..866U, 2014ApJ...783...44F}, we get that the cooling break energy is {\small $E_{e,br}=\frac{6\pi m^2_e}{\sigma_T}\,B^{-2}\,\tau^{-1}_{\rm age}$} and then the cooling synchrotron break energy is
\begin{equation}\label{synrad}
\epsilon^{\rm syn}_{\rm \gamma,br}=\frac{36\pi^2\,q_e\,m_e}{\sigma_T^2}\, B^{-3}\, \tau^{-2}_{\rm age}\,.
\end{equation}
The synchrotron spectrum is computed by summing over the electron distribution. The photon energy radiated in the range $\epsilon_\gamma$ to $\epsilon_\gamma + d\epsilon_\gamma$ is given by electrons with energies between $E_e$ and $E_e + dE_e$, then the photon spectrum can be estimated through the emissivity $\left[\epsilon_\gamma F(\epsilon_\gamma) d\epsilon_\gamma\right]_{\gamma, syn}=(- dE_e/dt)\,N_e(E_e)dE_e$,  with $F(\epsilon_\gamma)=\left[\frac{dN(\epsilon_\gamma)}{d\epsilon_\gamma}\right]_{\gamma, syn}$ the differential synchrotron photon flux. 
\subsection{Inverse Compton scattering}
Electrons accelerated at shock fronts can upscatter external photons up to higher energies. The synchrotron and inverse Compton scattering (IC) fluxes are  given by \citep{1997MNRAS.291..162A}
\be\label{rate}
\left[\epsilon^2_\gamma F(\epsilon_\gamma)\right]_{\gamma, IC}=\frac{U_{\rm syn}}{U_B}\,\left[\epsilon^2_\gamma F(\epsilon_\gamma)\right]_{\gamma, syn}\,,
\ee
with $F(\epsilon_\gamma)=\left[\frac{dN(\epsilon_\gamma)}{d\epsilon_\gamma}\right]_{\gamma, IC}$  the differential Compton scattering photon flux, and  $U_{syn}$ and $U_B=\frac{B^2}{8\pi}$ are the energy density of synchrotron radiation and magnetic field, respectively.   Considering the energy density of the CMB radiation \citep{1997MNRAS.291..162A}, the fluxes of synchrotron radiation at radio wavelengths and inverse Compton scattering in $\gamma$-rays are related by
\be\label{rate1}
\left[\epsilon^2_\gamma N(\epsilon_\gamma)\right]_{\gamma, IC}=\frac{U_{\rm CMB}}{U_B}\,\left[\epsilon^2_\gamma N(\epsilon_\gamma)\right]_{\gamma, syn}\,,
\ee
with $U_{CMB}=0.25\, {\rm eV\, cm^{-3}}$.
\subsection{Nonthermal Bremsstrahlung}
The Coulomb energy-loss rate of relativistic electrons, {\small$dE/dt \propto 3\,n_e\,\sigma_T\,m_e\,v^{-1} \ln \Lambda/2$} \citep{1979ApJ...227..364R}, is proportional to the gas density $n_e$ and independent of the electron energy, where $v$ is the velocity of nonthermal electrons and $\Lambda$ the Coulomb logarithm.  Comparing the cooling time of electrons due to Coulomb interactions and the age of the object $\tau_{age}$, we get that the Coulomb break kinetic energy is  \citep{2002ApJ...571..866U}
\be\label{Ecou}
K_{\mbox{\tiny Cou}}=\frac32 \sigma_T\,m_e\,n_e\,v^{-1} \ln \Lambda\,\tau_{age}\,.
\ee 
The differential energy spectrum of the bremsstrahlung emission from accelerated electrons is given by \citep{1970RvMP...42..237B}
\bary
\left[F(\epsilon_\gamma)\right]_{\gamma,Cou}&\simeq& \int dE_e\,N(E)\,\beta\,\cr
&&\hspace{0.1cm}\times \left(n_p\frac{d\sigma_{eH}}{d\epsilon}+ n_{He}\frac{d\sigma_{eHe}}{d\epsilon}+n_e\frac{d\sigma_{ee}}{d\epsilon}  \right)\,,\cr
\eary
where $n_p$, $n_{He}$ and $n_e$ are the hydrogen, helium, and electron number densities, respectively, $d\sigma/d\epsilon$ is the differential cross section for emitting a bremsstrahlung photon in the energy interval $\epsilon$ to $\epsilon+d\epsilon$ and $F(\epsilon_\gamma)=\left[\frac{dN(\epsilon_\gamma)}{d\epsilon_\gamma}\right]_{\gamma, Cou}$ is the differential bremsstrahlung photon flux.  
\subsection{Proton-proton Interactions}
High-energy protons accelerated in shock fronts can interact with ambient protons producing neutral and charged pions and mesons  \citep{1973ApJ...185..499S}.  For the accelerated cosmic ray distribution we adopt several possibilities including a standard power-law in momentum, $p$, $\frac{dN_p}{dp} = K_p\, p^{-\alpha_p}$, a power-law with an exponential cutoff, $\frac{dN_p}{dp} = K_p\, p^{-\alpha_p}e^{-\frac{p}{p_c}}$, and a broken power-law distribution, given by
\be\label{pr_dist}
\frac{dN_p}{dp}= K_p\left[1+ \left(\frac{p}{p_{br}}\right)^2 \right]^{-\frac{\Delta \alpha_p}{2}}\,\left(\frac{p}{p_0}\right)^{-\alpha_p}
\ee
with $\Delta \alpha_p=\beta_p-\alpha_p$.   In these functions $K_p$ is a normalization constant, $p_0$ is a scale constant, $p_{c}$ is the cut-off momentum and $p_{br}$ is the break momentum. We denote the corresponding cut-off and break energies in the $\gamma$-ray spectrum as $\epsilon^{\rm pp}_{\rm \gamma,c}$ and $\epsilon^{\rm pp}_{\rm \gamma,br}$.  Evidence in favor of this kind of proton distributions is seen in systems where SNRs interact with ambient material \cite[e.g.,][]{2009ApJ...706L...1A}.\\ 
If the proton distribution is converted to the total particle energy ($E_p$) space by $\frac{dN_p}{dE_p}=\frac{dN_p}{dp}\frac{dp}{dE_p}=\frac{E_p}{p}\frac{dN_p}{dp}$, the total energy content in protons can be calculated as 
\be
U=V\, W_p=V\,\int^{E_{p,max}}_{E_{p,min}}\,E_p\,\frac{dN_p}{dE_p} dE_p\,,
\ee
where the  particles are assumed to be uniformly distributed in a volume $V=\frac43 \pi R^3$ with radius $R$. The total hydrogen mass inside this volume is $M_H=n_p\, m_p\,V$. 
\subsubsection{GeV-TeV $\gamma$-ray spectrum}
The $\gamma$-ray spectrum produced by hadronic collisions is given by \citep{2006PhRvD..74c4018K}
{\small
\be\label{ph_inter}
\left(\frac{dN_\gamma}{dE_\gamma}\right)= \frac{V}{4\pi d^2_z}   n_p\int^\infty_{E_\gamma}\sigma_{pp}(E_p)\, \frac{dN_p}{dE_p}\,F^{pp}_\gamma \left(E_p, x\right)\frac{dE_p}{E_p}\,,
\ee
}
where $d_z=1.5\, {\rm kpc}$ is the distance to VER J2019+407,  $\sigma_{pp}(E_p)=34.3+1.88\, {\rm S}+0.25\,{\rm S^2}\,\,{\rm mb}$ is the cross section for these interactions, ${\rm S}=\log(E_p/{\rm 1 TeV})$ and $F^{pp}_\gamma(E_p,x)$ is given in \citet{2006PhRvD..74c4018K}.   It is noted that at high energies the gamma-ray spectrum maps the particle spectrum which could in principle be used to discriminate between different spectral functions.
%
%
%
%
%
%
%
\subsubsection{Neutrino emission}
Charged pions produced in hadronic collisions decay into electrons/positrons and neutrinos $\pi^{\pm}\rightarrow \mu^{\pm}+ \nu_{\mu}/\bar{\nu}_{\mu} \rightarrow e^{\pm}+\nu_{\mu}/\bar{\nu}_{\mu}+\bar{\nu}_{\mu}/\nu_{\mu}+\nu_{e}/\bar{\nu}_{e}$.  The spectrum of muon neutrinos generated by these interactions is given by\citep{2006PhRvD..74c4018K}
{\small
\be\label{ph_inter}
\left(\frac{dN_\nu}{d\en}\right)=\frac{V}{4\pi d^2_z}n_p\int^\infty_{E_\nu}\sigma_{pp}\, \frac{dN_p}{dE_p}\,F^{pp}_{\nu_\mu} \left(E_p, x\right)\frac{dE_p}{E_p}\,,
\ee
}
where the function $F^{pp}_{\nu_\mu}$ is split in two parts \citep{2006PhRvD..74c4018K}: one coming from  the decay of muons $\mu\to e\bar{\nu}_e\nu_\mu$ ($F^{pp}_{\nu^{(2)}_\mu}$) and the pion decay $\pi\to\mu\nu_\mu$ ($F^{pp}_{\nu^{(1)}_\mu}$).
\subsubsection{Neutrino Expectation}
The expected number of reconstructed neutrino events  in the IceCube experiment can be computed as \citep{2014MNRAS.441.1209F, 2014MNRAS.437.2187F}
\be
N_{ev}=T \,N_A\, \int_{\en^{th}} \sigma_{\nu N}(\en) \,M_{eff}(\en)\,  \frac{dN_\nu}{d\en}\,d\en\,,
\label{evtrate}
\ee
where $T\simeq$ 4 years is the observation time,   N$_A$=6.022$\times$ 10$^{23}$ g$^{-1}$ is the Avogadro number, $\sigma_{\nu N}(\en)=6.78\times 10^{-35}{\rm cm^2}(\en/TeV)^{0.363}$  is  the neutrino-nucleon cross section \citep{1998PhRvD..58i3009G},  $M_{eff}(\en)$ is the effective target mass of the IceCube experiment \citep{2013Sci...342E...1I} and the neutrino spectrum $dN_\nu/d\en$ is computed from hadronic interactions \citep{2015APh....71....1F, 2015APh....70...54F}.   The best function that reproduces the effective target mass is given by \citep{2016JHEAp...9...25F}
{\footnotesize
\bary\label{Veff}
M_{eff} =
\cases{ 
 f_6(E_\nu)\,\,, & $1\, {\rm TeV} < E_\nu < 950\, {\rm TeV}$ \nonumber \cr
 5.19\times 10^{-3}\left(\frac{E_\nu}{TeV}\right)+3.86\times 10^2,& $950 {\rm TeV} < E_\nu < 10^4 {\rm TeV}$,
 } 
\eary
}
where
{\footnotesize
\bary\nonumber
f_6(E_\nu)=2.46\times 10^{-15} \left(\frac{E_\nu}{TeV}\right)^6 -1.99\times 10^{-12} \left(\frac{E_\nu}{TeV}\right)^5\cr
-1.09\times 10^{-8} \left(\frac{E_\nu}{TeV}\right)^4\, +2.07\times 10^{-5} \left(\frac{E_\nu}{TeV}\right)^3\cr
 -1.38\times 10^{-2} \left(\frac{E_\nu}{TeV}\right)^2 +3.91 \left(\frac{E_\nu}{TeV}\right) -35.3\,.
\eary
}
It is worth noting that the muon neutrino of the effective target mass was used. \\
\section{Discussion}\label{discussion}
\subsection{Hadronic scenario}
\begin{figure}
\epsscale{1.5}
\plotone{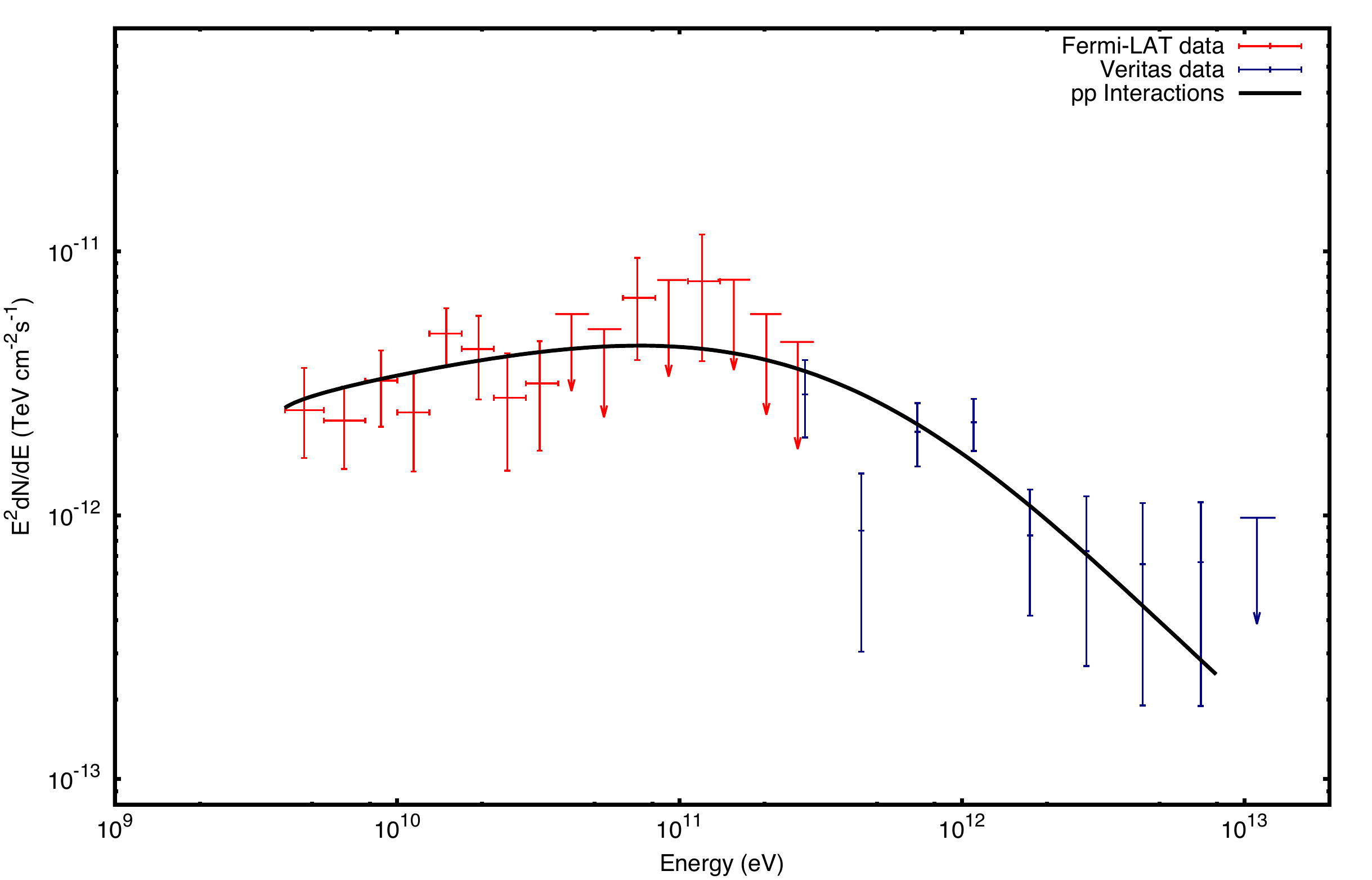}
\caption{GeV - TeV $\gamma$-ray SED of VER J2019+407. The solid line is the broken power-law function that best fits the Fermi-LAT fluxes (obtained in this work) and the VERITAS data, as shown in Table 1.\label{fig2}}
\end{figure}

\begin{figure}
\epsscale{1.5}
\plotone{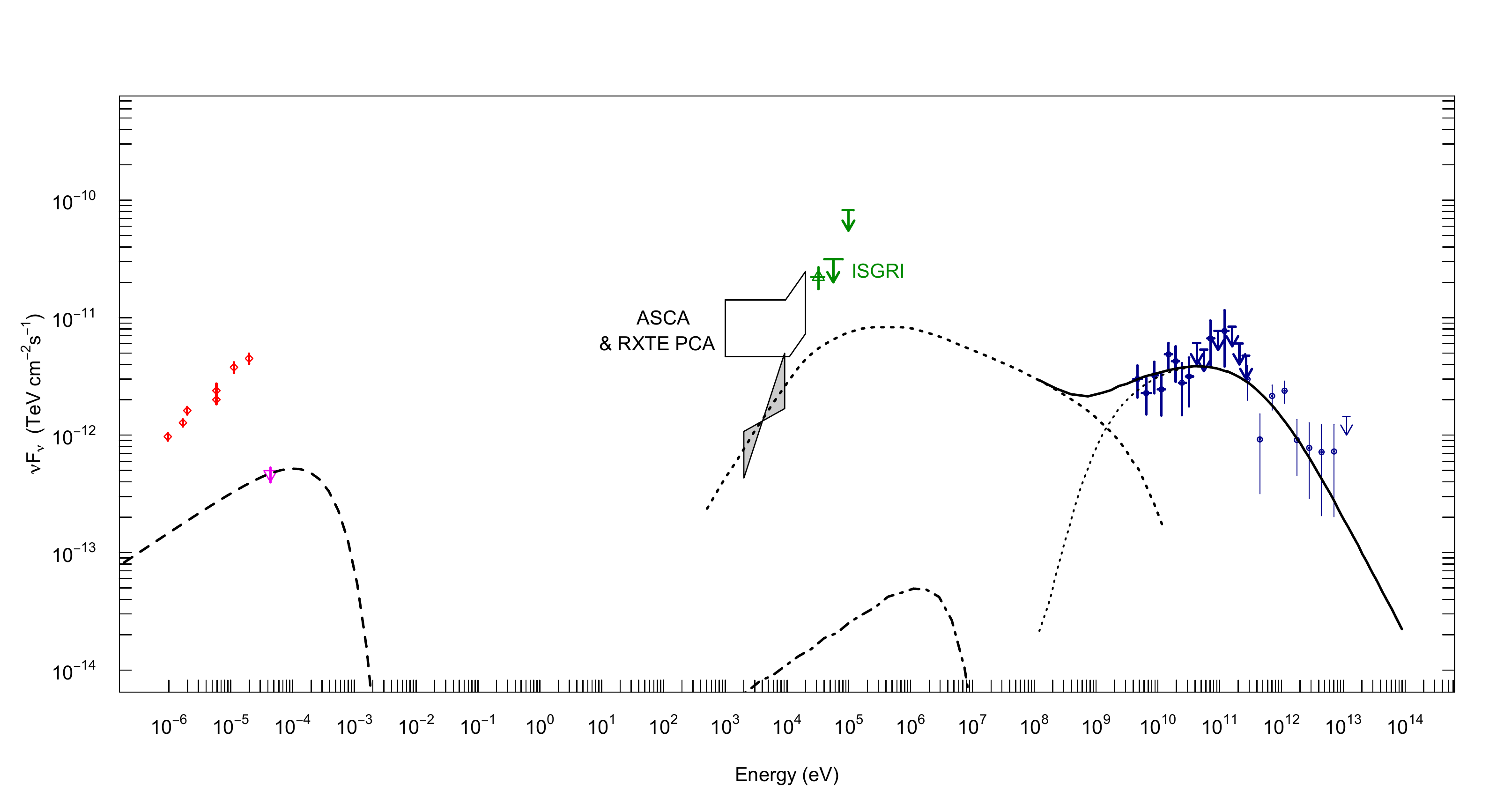}
\caption{Hadronic model for VER J2019+407. The radio SED points of the integrated emission from the SNR are shown in red while the magenta triangle shows a 10.6 GHz flux from a $7'\times9'$ region within VER J2019+407 \citep{1977AJ.....82..329H}. The ASCA and RXTE PCA observation of the remnant and an INTEGRAL-ISGRI observation of a clump to the west of VER J2019+407 are shown for reference \citep{2004A&A...427L..21B}. The shaded region corresponds to the ASCA observation of hard X-ray clumps seen in the region of TeV emission as published by Uchiyama et al. \citep{2002ApJ...571..866U}. The model components are synchrotron (dashed line), nonthermal bremsstrahlung (dotted line, ambient density $n_p=1$ cm$^{-3}$), IC-CMB (dash-dotted line), hadronic emission (thin dotted line, ambient density $n_p=1$ cm$^{-3}$) and the total (solid line).\label{fig3}}
\end{figure}

The GeV-TeV $\gamma$-ray SED of VER J2019+407 is shown in Fig. \ref{fig2}. The GeV Fermi-LAT fluxes are obtained in this work and the TeV data are taken from the VERITAS observation \citep{2013ApJ...770...93A}. As can be seen, the Fermi-LAT SED connects smoothly with the data at the highest energies. We present the results of fitting the $\gamma$-ray SED with a $\chi^2$ minimization as implemented in the ROOT software package \citep{1997NIMPA.389...81B} in Table 1 and the resulting parameters associated to each fitting function. We can see that a broken power-law best describes the overall GeV-TeV $\gamma$-ray SED and adopt the parent proton distribution given by eq. (\ref{pr_dist}) with $\Gamma_l=\alpha_p$ and $\Gamma_h=\beta_p$ to calculate the $\gamma$-ray emission in a hadronic scenario. The resulting total energy in cosmic ray hadrons in this model is $1.95\times10^{50}\,\left(\frac{n_p}{1\,{\rm cm}^{-3}}\right)^{-1}$ erg for $p_0=1\,$ TeV and a source distance of 1.5 kpc. The break momentum in the particle distribution is 0.45 TeV.\\  
\begin{center}
\scriptsize{\textbf{Table 1. Fitting results for the GeV-TeV spectrum of VER J2019+407.}}\\
\end{center}
\begin{center}\renewcommand{\tabcolsep}{0.3cm}
\renewcommand{\arraystretch}{1.5}
\begin{tabular}{lcccc}\hline\hline
{\small \bf Spectral shape} & {\small \bf Parameters}  & {\small \bf Values} & {\small \bf$\chi^2$/dof.}\\ \hline\hline
\multicolumn{1}{c}{\small Simple power-law} \\
\cline{1-1}
{\scriptsize Spectral index} & {\scriptsize $\Gamma$}  &  {\scriptsize  2.18$\pm$0.03} & {\scriptsize 1.38}\\ \hline\hline
\multicolumn{1}{c}{\small Power-law with cutoff} \\
\cline{1-1}
{\scriptsize Spectral index} & {\scriptsize $\Gamma$}     &  {\scriptsize  2.07$\pm$0.07} & {\scriptsize 1.20}\\
{\scriptsize  Cutoff energy}     & {\scriptsize $E_{\rm \gamma,c}$ (\rm TeV)}  &  {\scriptsize  3.55$\pm$2.39} & \\ \hline\hline
\multicolumn{1}{c}{\small Broken power-law$^a$} \\
\cline{1-1}
{\scriptsize Low spectral index}   & {\scriptsize $\Gamma_l$}     &  {\scriptsize  1.78$\pm$0.15} &  \\
{\scriptsize High spectral index}   & {\scriptsize $\Gamma_h$}  &  {\scriptsize  2.46$\pm$0.11} & {\scriptsize 1.02} \\
{\scriptsize Break energy}  & {\scriptsize $E_{\rm \gamma,br}$ {\rm (GeV)}}  &  {\scriptsize  70.7$\pm$0.1} &  \\ \hline\hline
\end{tabular}
\end{center}
\begin{flushleft}
\scriptsize{
$^a$ $\Gamma_l$ and $\Gamma_h$ are the spectral indices below and above the break energy $E_{\rm \gamma,br}$.\\
}
\end{flushleft}
The cosmic ray distribution below the break is harder than predicted for standard test-particle shock acceleration but could be accounted for by nonlinear effects such as shock modification by cosmic-rays \citep{1999ApJ...511L..53M, 1999ApJ...526..385B}. As discussed by \citet{2009ApJ...706L...1A}, the break in the proton spectrum could be related to the effects of damping of magnetohydrodynamic turbulence due to ion-neutral collisions in a zone of interaction of the shock with ambient material, assuming that this break is not intrinsic to the acceleration process and that acceleration takes place near the Bohm limit \citep{1999ApJ...513..311B}.\\
The leptonic emission in this model is calculated with a power-law electron distribution with an index of 2.3, a cutoff particle energy of 20 GeV and a magnetic field of 9.5 $\mu$G. The total lepton energy is $7.7\times10^{48}$ erg and the ambient density used to calculate the bremsstrahlung emission is the same as that used for the hadronic component, 1 cm$^{-3}$. The resulting SED and model is shown in Figure \ref{fig3}. We use a bremsstrahlung model similar to the one by \cite{2002ApJ...571..866U} to account for the hard X-ray emission. Although a steeper electron distribution could be used to lower the bremsstrahlung contribution at $\gamma$-ray energies we chose instead to place a cutoff in the electron distribution above a particle energy of 20 GeV.\\
Considering the resulting neutrino counterpart from pp interactions in this source (eq. \ref{pr_dist}), using the values found for the indices ($\alpha_p$ and $\beta_p$),  the total energy in the hadrons,   the break momentum in the particle distribution and the target mass of the IceCube experiment (eq. \ref{Veff}) we have computed the number of neutrinos expected in the IceCube detector, and we find that more than $10^3$ years of data taking are needed to detect one event at dozens of TeV.

\subsection{Leptonic scenario}
\begin{figure}
\epsscale{1.5}
\plotone{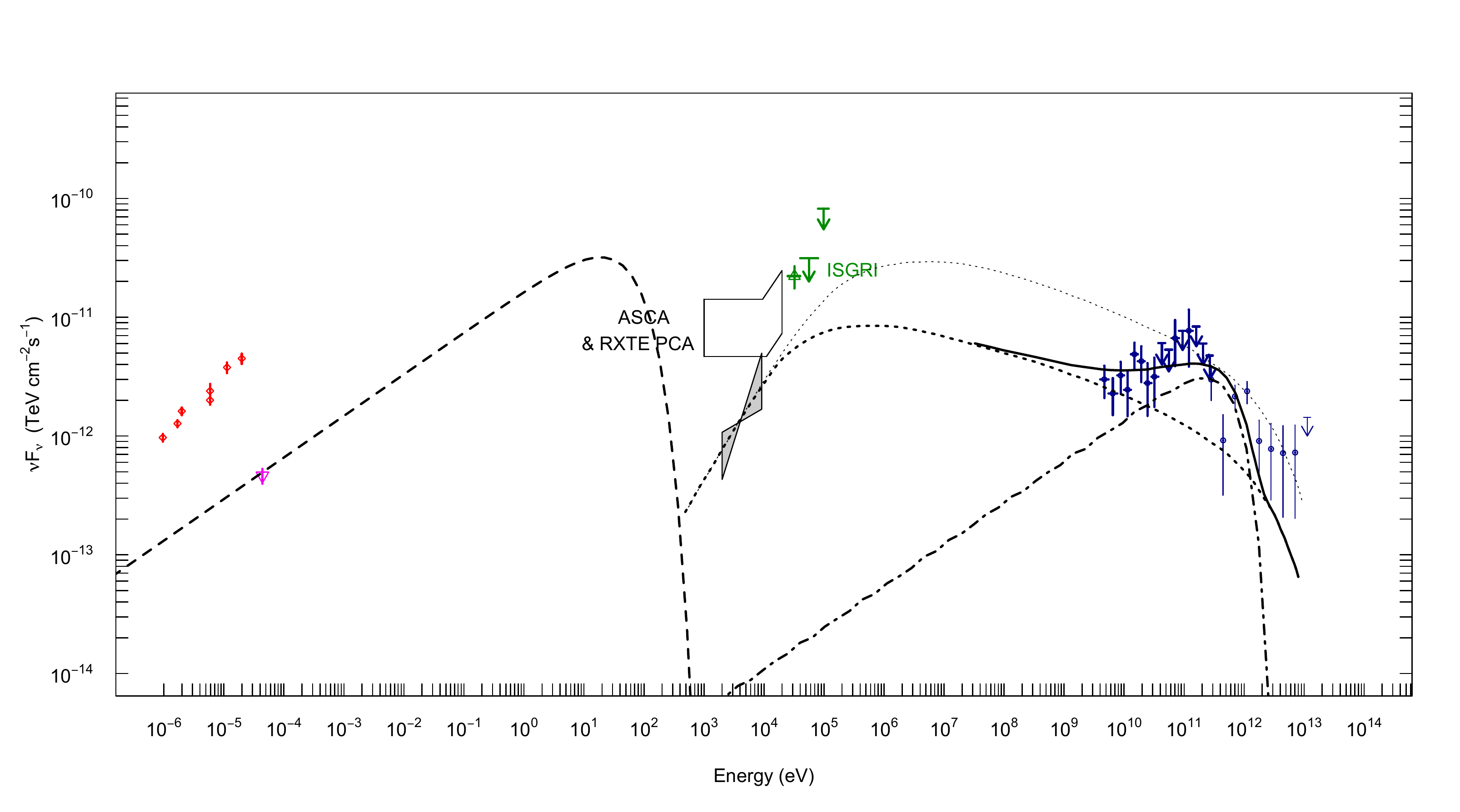}
\caption{Leptonic model for VER J2019+407. The data shown is the same as in Fig. \ref{fig3}. The model components are synchrotron (dashed line), nonthermal bremsstrahlung (dotted line, ambient density $n_p=1$ cm$^{-3}$), IC-CMB (dash-dotted line) and total gamma-ray emission (solid line). The thin dotted line shows the bremsstrahlung model by \cite{2002ApJ...571..866U} which they calculated for similar parameters except an ambient density of $n_p=10$ cm$^{-3}$ which overpredicts the measured fluxes.\label{fig5}}
\end{figure}

We find that a simple scenario where a population of HE electrons whose energy distribution is a power-law producing gamma-rays through IC scattering of CMB photons cannot reproduce the gamma-ray data for reasonable distribution index values. One of the main issues with this scenario is that it cannot account for the hard X-ray fluxes seen by ASCA. This has already been noted by \cite{2002ApJ...571..866U}. They ruled out synchrotron (and IC) as the origin of the X-ray emission and conclude that a plausible scenario to account for the observations is nonthermal bremsstrahlung from HE electrons as it would be expected from Coulomb interactions that give rise to a hard X-ray spectrum. We adopt similar parameters as these authors for the models such as a magnetic field of 9 $\mu$G, a particle spectral index of 2.3 and a cutoff electron energy of 10 TeV.\\
The radio flux from the electrons predicted by \cite{2002ApJ...571..866U} is about 10\% of the total flux from the SNR, which is in agreement with the observation at 10.6 GHz. However, their SED models overpredict the $\gamma$-ray fluxes. We then take their emission model for an electron distribution index of 2.3 and a gas density $ 10$ cm$^{-3}$ and scale it down for a lower density value of 1 cm$^{-3}$. We also estimate the kinetic electron energy below which the particle spectrum flattens by equating the source age (7000 yr) to the cooling time due to Coulomb interactions which gives the result $K_{\mbox{\tiny Cou}}\sim 0.22$ MeV, and note that the bremsstrahlung spectrum can be described by a broken power law, with the index above some photon energy in the hard X-ray regime equal to that of the particle spectral index and increasing by one power below this energy \citep{2002ApJ...571..866U}. The resulting model is shown in Figure \ref{fig5} together with the expected IC-CMB contribution. The total lepton energy is $7.7\times10^{48}$ erg for a magnetic field of 9 $\mu$G. This simple leptonic scenario is able to account for the $\gamma$-ray fluxes reasonably well although an improved model, which is beyond our main objective in this work, could consider distinct emitting zones with different spectral and ambient properties.
\section{Conclusion}\label{conclusion}
With analysis of Fermi LAT data we have found a region of enhanced hard GeV emission within the shell of SNR G78.2+2.1 which is in spatial coincidence with the TeV source VER J2019+407. The joint GeV-TeV spectrum can be best described with a broken power-law. We cannot rule out the leptonic nor the hadronic mechanism for the origin of the $\gamma$-rays.\\
A simple leptonic scenario with nonthermal bremsstrahlung and IC emission can account for the gamma-ray SED. Bremsstrahlung emission dominates at low $\gamma$-ray energies and is able to explain the hard X-ray emission seen by ASCA in clumps within the extent of VER J2019+407 with an average ambient density of 1 cm$^{-3}$ and a magnetic field of 9$\mu$G, for a total particle energy of $7.7\times10^{48}$ erg.\\
In the hadronic scenario, the required spectral index of the particle population is lower than the standard result from diffusive shock acceleration, but this could be explained by nonlinear effects \citep{1999ApJ...511L..53M, 1999ApJ...526..385B}. The origin of the spectral break is also of theoretical interest. The total energy in the particles is within expected values, $\sim 1.95\times10^{50}\,\left(\frac{n_p}{1\, {\rm cm^{-3}}}\right)^{-1}$ erg. In this scenario, the leptonic emission can be suppressed with a particle cutoff energy of 20 GeV. A considerable amount of nonthermal bremsstrahlung emission is expected around several hundred MeV if the hard X-rays are indeed associated to the HE electrons in the SNR. Detection of synchrotron emission from the northern shell above a frequency of 11 GHz could help quantify the maximum lepton energy and thus the amount of bremsstrahlung $\gamma$-ray emission. We point out that it is certainly possible to have a scenario with mixed bremsstrahlung and hadronic emissions depending, among other parameters, on the true electron power-law index in the northern shell.  We computed the neutrino flux resulting from pp interactions and also estimated the number of events expected in the IceCube detector.  We found that $\sim 10^3$ years of data taking are needed to observe one event at dozens of TeV. \\
From our analysis, there is indication that the southern shell of the SNR has a softer GeV spectrum, and indeed this region is not detected at TeV energies. Observations yielding the magnetic field values and its geometry or the presence of dense material such as molecular clouds and other ambient properties in the north of G78.2+2.1 could help explain wether the enhanced gamma-ray emission is due to an increase in acceleration efficiency or the result of abundant target material in the case of the hadronic scenario. A detailed analysis of the LAT emission below $\sim$4 GeV is required to better constrain the model parameters and the nature of the radiation. We carried out a preliminary analysis of LAT data above 1 GeV which indicates that the source spectrum becomes softer, as predicted by the models in the previous Section. However, we cannot say if this is an artifact from contamination by emission from the bright pulsar within the shell of the SNR.
\acknowledgments
We thank the anonymous referee for a critical reading of the paper and valuable suggestions that helped improve the quality and clarity of this work. We are also grateful to PAPIIT-UNAM IG100414 and Universidad de Costa Rica for financial support. This research has made use of NASA's Astrophysical Data System and the Canadian Galactic Plane Survey (CGPS) which is a Canadian project with international partners. The Dominion Radio Astrophysical Observatory is operated as a national facility by the National Research Council of Canada. The CGPS is supported by a grant from the Natural Sciences and Engineering Research Council of Canada.
%
%

%
%
\end{document}